# Artificial Intelligence as an Anti-Corruption Tool (AI-ACT) Potentials and Pitfalls for Top-down and Bottom-up Approaches

Nils Köbis[1], Christopher Starke[2,3] & Iyad Rahwan[1]

[1] Center for Humans and Machines, Max-Planck-Institute for Human Development

[2] Heinrich Heine University Düsseldorf

[3] Heine Center for Artificial Intelligence and Data Science

Correspondence to: Nils Köbis, koebis@mpib-berlin.mpg.de, Max-Planck-Institute for Human Development, Center for Humans and Machines, Lentzallee 94, 14195 Berlin, Tel.: +49 30 82406-751/752

**Abstract**

Corruption continues to be one of the biggest societal challenges of our time. New hope is placed in Artificial Intelligence (AI) to serve as an unbiased anti-corruption agent. Ever more available (open) government data paired with unprecedented performance of such algorithms render AI the next frontier in anti-corruption. Summarizing existing efforts to use AI-based anti-corruption tools (AI-ACT), we introduce a conceptual framework to advance research and policy. It outlines why AI presents a unique tool for *top-down* and *bottom-up* anti-corruption approaches. For both approaches, we outline in detail how AI-ACT present different potentials and pitfalls for (a) input data, (b) algorithmic design, and (c) institutional implementation. Finally, we venture a look into the future and flesh out key questions that need to be addressed to develop AI-ACT while considering citizens' views, hence putting "society in the loop".

Corruption – commonly defined as the abuse of entrusted power for private gains – presents one of the biggest societal and political challenges of our time (Fisman and Golden, 2017a). Emerging in various forms, it undermines efficiencies of public institutions, widens gaps in inequalities and thereby hinders the achievement of the UN sustainable development goals (Rothstein and Varraich, 2017; Mungiu-Pippidi and Heywood, 2020). The ongoing COVID-19 pandemic has further exacerbated this trend, confirming previous research showing that corruption augments the adverse effects of catastrophes (Ambraseys and Bilham, 2011). While vast (financial) efforts have been invested in the fight against corruption, they have shown little signs of success (Fisman and Golden, 2017b; Mungiu-Pippidi, 2017).

The advancement of digital technologies (Mattoni, 2020), in particular artificial intelligence (AI; Aarvik, 2019), provide new hope to fight corruption more effectively. AI, defined by the EU High-Level Expert Group on Artificial Intelligence as "systems that display intelligent behaviour by analysing their environment and taking actions – with some degree of autonomy – to achieve specific goals" (Ala-Pietilä *et al.*, 2019) differs from static information and communication technologies (ICTs). Namely, "classic" ICTs enable digitalization of procurement procedures, provide public services online and publish open government data (Makowski, 2017; Adam and Fazekas, 2018; Kossow, 2020). Yet, classic ICTs do not act autonomously.

This is where AI goes a step further. Thanks to its learning abilities, AI can autonomously execute a wide range of tasks previously reserved to human actors (Rahwan *et al.*, 2019). Already implemented in several pioneering projects, it can take over anti-corruption tasks like predicting, detecting and disclosing corruption cases (López-Iturriaga and Sanz, 2018; Martins, Guimaraes and dos Santos, 2018; Lima and Delen, 2020). Yet, in contrast to other applications of AI to fight crime (Ryman-Tubb, Krause and Garn, 2018; Crawford, 2019), AI for anti-corruption is not a tool for governments to scrutinize its citizens, but can be a tool for citizens to scrutinize their government. Thus, AI in the fight against corruption evokes much enthusiasm – being praised "as the next frontier in anti-corruption" (Petheram, 2018).

While first collections of cases using AI in the fight against corruption exist (Aarvik, 2019), research mapping the potential and pitfalls of using AI anti-corruption tools (henceforth AI-ACT) is lacking. Filling this gap, we introduce a conceptual framework that systematically introduces AI to the rich literature on (anti-)corruption (Hough, 2013; for example Hardi, Heywood and Torsello, 2015; OECD, 2017; Mungiu-Pippidi and Heywood, 2020). In this work, we outline why AI presents a unique tool for *top-down* and *bottom-up* anti-corruption approaches. For both approaches, we outline in detail how harnessing the potential of AI-ACT depends on (a) input data, (b) algorithmic design, and (c) institutional implementation. Finally, we venture a look into the future and flesh out key questions that need to be addressed to develop AI-ACT while considering citizens' views, hence putting "society in the loop" (Rahwan, 2018).

**How to harness the potential of AI-ACT**

To grasp how AI can help curbing corruption requires a closer look at corruption and AI. Recent corruption research has put forth a plethora of frameworks differentiating between various forms of corruption (Lange, 2008; Heidenheimer and Johnston, 2011; Marquette and Peiffer, 2015; Bauhr, 2017; Köbis and Huss, 2018). In fact, corruption is widely understood as an umbrella term that encompasses many different behaviors (Bussell, 2013; Köbis *et al.*, 2016). Current directions in corruption research outline the importance of specifying the respective type of corruption one seeks to combat before embarking on an anti-corruption crusade (Kubbe and Engelbert, 2017; Mungiu-Pippidi and Heywood, 2020). For example, tackling embezzlement within the political sector differs profoundly from changing bribery schemes in the police force. We agree with this notion. At the same time, since we face the initial phase of AI introduction to anti-corruption and because many of the described effects below are not corruption-type specific, here we first outline the general interplay of AI and anti-corruption.

Such AI-based anti-corruption tools are primarily fueled by recent breakthrough advances in machine learning (Rahwan *et al.*, 2019). Such technologies already match, or even exceed, human abilities across a growing range of (narrow) tasks, spurring hope that AI can contribute to more effective anti-corruption efforts (Aarvik, 2019). Overall, interest in AI is on the rise (again). Ever more available data and surging computing power enable AI technologies to achieve major breakthroughs (Bostrom, 2017; Tegmark, 2017), for example, in the domain of Natural Language Processing (NLP; Radford *et al.*, 2019; Torabi Asr and Taboada, 2019; Köbis and Mossink, 2021). The uniqueness of AI compared to other technologies lies in its autonomous, learning abilities. Instead of a programmer specifying the machine's course of action for all possible outcomes, AI algorithms can figure out solutions themselves, some of which are unpredictable, even for the human programmers (Silver *et al.*, 2017). Many hope that these autonomous learning abilities can contribute to the increasingly data-driven efforts to fight corruption (Adam and Fazekas, 2018; Mattoni, 2020).

This hope receives traction as public administration becomes increasingly digital (Starke, Naab and Scherer, 2016; Kossow, 2020). In pursuit of more transparency, e-government initiatives, open data programs and citizen-driven crowdsourcing efforts are making ever more data publicly available (Attard *et al.*, 2015; Mayernik, 2017). According to a long-standing assumption, this growing availability of information will enable citizens to educate and coordinate themselves in the fight against corruption (Schroth and Sharma, 2003). However, recent policy-oriented research reveals that the mere disclosure of data does not suffice to curb corruption (Murillo, 2015). Someone, whether these are prosecutors, journalists or civil society actors need to be able to draw inferences from data to render it actionable for policy efforts (Jasanoff, 2017). Put differently, autonomous agents are needed to put *transparency in action* to advance accountability. Without this resource-intensive engagement, the mere availability of such data remains *transparency on paper*. Succinctly put, transparency without accountability is like the "sound of one hand clapping" (Seligsohn, Liu and Zhang, 2018, p. 804).

AI brings new potential to put *transparency in action*. It saves human resources by taking over key tasks of (pre-)screening large datasets, analyzing it to enable detecting, predicting and reporting risks, suspicions or clear-cut cases of crimes (Metz and Satariano, 2020), like corruption (Grace *et al.*, 2016). While in principle humans could engage in these tasks, the sheer magnitude of many Big Data sources renders this task most often infeasible. Supercharged by ever more available data and increasing computing power, AI algorithms can draw inferences by analyzing, modeling and dynamically visualizing these data (Verhulst, Engin and Crowcroft, 2019). Moreover, its autonomous abilities allow AI technologies to act based on the insights gained, without requiring further engagement of human actors, like automatically disclosing suspicious cases of corruption (Mattoni, 2020). In short, AI can make autonomous decisions about corruption (risks) based on Big Data.

These autonomous abilities of AI can be used for anti-corruption efforts adopting top-down or bottom-up approaches. On the one hand, *top-down* approaches rest on the view that institutions are shaped by laws crafted by political leaders (Easterly, 2008). Hence, top-down anti-corruption efforts seek to bring about change by introducing new laws, regulations and procedures within public administration. AI can be used to assist such approaches. Consider for example Kalsada, a project in the Philippines using ML to map roads and assess their quality to identify possible cases of embezzlement of road building material (Kaiser, Rahemtulla and Van den Brink, 2016). Here, AI can drastically improve efficiency compared to previous hands-on approaches that required to literally dig up the road to assess the quality of the materials used (Olken, 2007). Hence, AI-ACT in top-down approaches provide a powerful new tool to those tasked with anti-corruption, such as prosecutors and compliance officers.

*Bottom-up* approaches, on the other hand, view institutions as emerging rather spontaneously through social norms, customs, traditions, beliefs and values within a society (Easterly, 2008). Bottom-up anti-corruption efforts seek to analyze the given cultural and societal context and then identify as well as support existing efforts within the society that seek to reduce corrupt practices (Khan, Andreoni and Roy, 2016). This approach chiefly relies on active civil society organizations and journalists who can play a watchdog role (Camaj, 2013; Starke, Naab and Scherer, 2016; Köbis, Iragorri-Carter and Starke, 2018). AI can assist such bottom-up approaches, such as in the case of the Ukrainian portal Dozorro that draws on AI to flag public procurement tenders with high risk of corruption and communicates them to the public (Oksha, 2019).

For both approaches, using AI to fight corruption differs from other efforts in which AI is used to fight other forms of crime. That is, when tackling other forms of misconduct, the government uses AI tools to inspect *its citizens*. Yet, AI-ACT present tools for governments (top-down) and, more importantly, citizens (bottom-up) to *inspect the government* (officials). Consider that much warranted attention is already placed on AI tools for fraud detection (Ryman-Tubb, Krause and Garn, 2018), predictive policing (Richardson, Schultz and Crawford, 2019) and pretrial risk assessment, such as the by now famous COMPAS algorithm (Angwin *et al.*, 2016; Dressel and Farid, 2018). In these efforts the government, often in liaison with private corporations, uses AI to police its citizenry (Crawford *et al.*, 2019), causing several challenges (Alston, 2019). The unique twist of AI-ACT is that they are tools to tackle abuse of power by those in government. Instead of the government taking the role of big brother watching over the citizens, AI-ACT, in particular when used by bottom-up efforts, allow the public to turn into many little watchdogs keeping the government in check.

These relatively new forms of using AI to fight corruption have received little empirical attention. Hence, we outline in more detail the unique potentials and pitfalls for AI-ACT in top-down and bottom-up approaches along the factors of *input data*, *algorithmic design* and *institutional implementation* (see Table 1).

**Table 1.** Overview of the emerging potential and pitfalls of using AI-ACT for top-down and bottom-up approaches.

| Artificial Intelligence-based Anti-Corruption Tools (AI-ACT) | | |
|---|---|---|
| | **Top-Down Approaches** | **Bottom-Up Approaches** |
| **Actors** | Criminal investigators, prosecutors, compliance officers, auditors | Journalists, bloggers, civil-society activists, average citizens |
| **Input data** | Classified & open government data, crowdsourced data, (social) media text | Open government data, data leaks, crowdsourced data, (social) media text |
| **Algorithmic design** | Rather minimize false-negative rate | Rather minimize false-positive rate |
| **Institutional Implementation** | Human-out-of-the-loop to escape the corruption trap | Human-in-the-loop to ensure legitimacy |
| **Examples** | MARA, Kalsada | Rosie da Serenata, Botivist, Dozorro |

**Potentials and Pitfalls of AI-ACT**

**Input Data**

Powered by Big Data, the success of AI-ACT crucially depend on the quality and size of available input data. A first key question pertains to the *source of the data*. Current AI-ACT draw on three main data sources: First, (*open*) *government data* describe databases containing information collected and provided by the government such as data about internal contracts and procedures. The emergence of e-government and open government initiatives is rendering many such data sources publicly available (Adam and Fazekas, 2018; Dávid-Barrett and Fazekas, 2020). One example are the pioneering efforts in Ukraine releasing information about public tenders via the portal of ProZorro (Huss and Nesterenko, 2016). However, large chunks of government data frequently remain undisclosed, unless hackers or whistleblowers expose them, which brings us to the second data source: *data leaks.* Several high profile cases of data leaks, such as the Panama Papers or more recently the FinCen Files have helped to unveil impactful corrupt cases (Baack, 2016; Köbis and Starke, 2017; Obermaier and Obermayer, 2017; Leopold, 2020).

Third, *crowdsourced data* describes efforts by citizens to expose corruption (Mattoni, 2020). One of the most famous examples is the Indian crowdsourcing portal Ipaidabribe.com (Ang, 2014). On the website, citizens can report cases in which they (were forced to) pay a bribe, tallying up to more than 197,000 of such reports. Relatedly, the fourth main data source consists of *(social) media text.* Advances in NLP have enabled large-scale analyses of such text data sources, also in the fight against corruption. For one, existing AI-ACT in Spain make use of classical news media reports of corruption to predict future occurrences of corruption on a regional level (López-Iturriaga and Sanz, 2018). Furthermore, one project called Botivist entails a Tweetbot that automatically contacts those who tweet about corruption, encouraging them to engage in collective action against corruption, such as signing petitions (Savage, Monroy-Hernandez and Höllerer, 2016).

The different types of data sources present different issues of *accessibility* and *ethics* for top-down and bottom-up approaches. When it comes to *accessibility*, top-down efforts using AI-ACT can frequently draw on internal and undisclosed government data (Grace *et al.*, 2016). Consider for example the Brazilian Department of Research and Strategic Information (DIE) at the Brazilian Office of the Comptroller General (CGU) that calculates an individual-level corruption score by training an ML algorithm on undisclosed government data (Carvalho, Carvalho and Ladeira, 2014). Although restrictions also apply for such top-down efforts, overall top-down efforts have more access compared to bottom-up efforts. That is because bottom-up efforts cannot draw on undisclosed government data, unless it becomes openly available - either voluntarily via open government initiatives or involuntarily via data leaks and crowdsourcing.

The modalities of disclosure influence *ethical* issues that can arise, which are seldom when using open government data but more common using involuntarily disclosed data. Let us first consider data leaks that can raise privacy concerns about the information disclosed. Respecting individuals' privacy while still being able to draw meaningful inferences about corruption represents a challenge. While responsible investigative journalists often wait to publicize data until sensitive information is sufficiently masked, unmasked disclosures, such as some of the data leaked on Wikileaks, neglect such privacy concerns (Cerulus, 2016). Violating privacy rights, even of potential corrupt culprits, does not only jeopardize the legality of using such data but also undermines the perceived legitimacy of such anti-corruption efforts.

Ethical issues also arise when using crowdsourced data. Consider reporting portals like Ipaidabribe. Here, one main problem arises if the often-anonymous reports can inaccurately denunciate individuals holding public office. In an attempt to meet this challenge, Ipaidabribe disables people to personally identify culprits, hence somewhat reducing incentives for wrongful accusations (Hough, 2015).

The second main issue for AI-ACT relates to the *quality of the data*, captured by the notion of "garbage in, garbage out" (Salganik, 2018). As with all kinds of data, large amounts of digital data also need to be carefully evaluated in terms of whether (1) they entail good proxies for corruption (i.e. validity) and (2) they are a consistent representation of the underlying factor of the measure (i.e. reliability). Two contrasting examples help to illustrate this point. On the one hand, crowdsourced data via Ipaidabride are valid measures of petty corruption in that they represent citizens' everyday experiences with (corrupt) public officials. However, each data point may lack reliability, as the veracity of a given report cannot be ensured. On the other hand, existing AI tools already calculate "Crime and Corruption Risk Scores", also drawing on highly contested and unreliable data sources such as facial images (Wu and Zhang, 2016). The issue of data quality becomes particularly important in the domain of corruption as obtaining valid and reliable data about corruption (in machine learning referred to as *ground truth*) presents a thorny challenge in the first place (Schwickerath, Varraich and Smith, 2016).

Another aspect of data quality are systematic *biases*. Although algorithms have the aura of being impartial, objective and hence fair, by now ample empirical evidence from various domains documents that ML algorithms can suffer from systematic biases (Kleinberg *et al.*, 2018; Crawford *et al.*, 2019; Obermeyer *et al.*, 2019). Algorithms trained on biased data sets, reproduce and further exacerbate the existing biases in the society (Veale and Binns, 2017). Consider the Brazilian project MARA that calculates an individual-level corruption score based with an algorithm trained on previous conviction data (Marzagão, 2017). If the data used to train the algorithm suffers from biases, such as minority groups being prosecuted and charged more frequently, algorithmic predictions will be skewed and reproduce these biases.

Taken together, AI-ACT are only as good as their input data. Having access to useful data presents the first challenge, one where top-down efforts have an overall advantage over bottom-up efforts that rely on (voluntarily or involuntarily) released data. Moreover, once granted access, the usefulness of predictions by AI-ACT hinge on the quality of the data - both in terms of whether they contain a good proxy for corruption and how biased the data are. With input data at hand, the next step in the process pertains to the technical calibration of the algorithm.

**Algorithmic Design**

Encapsulated in the notion that AI is an ideology, not a technology (Lanier and Weyl, 2020), design choices about how to program a given algorithm often have far-reaching, value-laden consequences. For AI-ACT, the technical requirements to properly calibrate the algorithm itself differ starkly along top-down and bottom-up approaches. A first key feature to consider in the algorithmic design is the *accuracy* of the predictions. Here, a trade-off occurs between false positive errors and false negative errors (Harrison *et al.*, 2020). False positive errors describe the wrong classification of innocent individuals as "corrupt". Since corruption accusations come with a strong stigma, such errors bear a large cost for those wrongfully accused.

Conversely, false negatives come with the cost of leaving actual corrupt cases undetected, which means a non-negligible cost for public institutions or society as a whole (Fisman and Golden, 2017a). Obviously, algorithms with an overall high accuracy rate are preferable over those with low predictive accuracy. However, oftentimes, a decrease in one type of error comes with an increase in the other error type. Consequently, challenging trade-offs emerge when calibrating the algorithm, such as deciding which variables to include in the AI model and which type of errors to reduce (Kearns and Roth, 2019).

These trade-offs differ across top-down and bottom-up approaches because errors cause asymmetric costs. For top-down approaches, the decisions made by AI-ACT are mostly communicated internally, for example to a compliance officer. Ideally, that person then further checks the validity of the flagged suspicion and, upon confirmation, decides about which actions to take next. Due to this internal checks and balances mechanism, the risk of losing (public) reputation by being wrongly accused is greatly alleviated. Thereby, for top-down efforts, it is arguably more important to calibrate algorithms towards minimizing the false negative rate (rather than the false positive rate), to avoid that corrupt offenses remain undetected.

For bottom-up approaches, however, such checks and balances mechanisms are typically lacking by directly publicizing cases of suspicion. False positive errors of corruption accusations come with the immense cost of a stigma that is often hard to shake, even after the accusations turn out to be false (Rauh, 2018). First cases of wrongful crime accusations based on faulty facial recognition algorithms have been documented (Hill, 2020). Following such false positive errors, those accused of corruption tend to be prematurely prosecuted in the court of public opinion, and suffer irreversible reputation losses (Rauh, 2018). Therefore, when bottom-up approaches make use of AI-ACT reducing false positive errors arguably presents a more pressing priority than reducing false negative errors.

Taken together, calibrating accuracy of AI-ACT can play a vital role in the success of such efforts. In general, when top-down approaches utilize algorithms they can afford to over-report corruption suspicions. Yet, when using AI-ACT for bottom-up approaches the reputation of the accused is at stake, hence calibrating such algorithms to be more lenient is advisable.

**Institutional Implementation**

Algorithms never operate in a vacuum, but are embedded in social contexts (Berendt, 2019). Obviously, a wide range of choices accompanies the complex process of implementing AI-ACT. It starts with the question whether or not an AI-ACT should be implemented in the first place. Those seeking to use AI to fight corruption need to critically examine whether AI tools can in fact help to reduce a given corruption scheme or whether AI technology is merely applied due to its novel, edgy allure. At times, implementing a static ICT or introducing a simple regression prediction model might do the trick.

Furthermore, when implementing AI-ACT in a top-down manner it is important to be aware of the immense importance of trust towards such technologies (Kennedy, Waggoner and Ward, 2018). People quickly distrust new technologies if the decision about the implementation happens behind closed doors or if the workings of the algorithms themselves remain hidden from public scrutiny (Metz and Satariano, 2020). Among public officials, hence those subjected to the algorithmic predictions, such an intransparent implementation can elicit a sense of AI surveillance (Möhlmann and Zalmanson, 2017). As one consequence, tech backlash can rapidly emerge. For instance, using sensitive data to prevent corruption and the accompanying feeling of being under constant surveillance may have adverse effects on public officials. They might start shirking, i.e. work less, and lose trust towards their employee, which can eventually lead to an exodus of the talented workforce. In specific contexts, such negative effects may cause more harm than is justified by the benefit of catching a few more "corrupt culprits".

Also, bottom-up approaches that use algorithms without careful implementation can similarly backfire. One concrete threat consists of "spamming" citizens with corruption cases, irrespective of whether they are true or not. False accusations disseminated by AI-ACT can have a desensitizing effect on citizens making them care less about the issue. But, also true accusations can have inadvertent ramifications. For instance, empirical evidence suggests that continuous exposure to negative political news can foster cynicism, (the so-called 'media malaise hypothesis', Marcinkowski and Starke, 2018). Citizens who learn about widespread corruption within political elites may turn their back on the anti-corruption efforts and potentially even feel licensed to break ethical rules for their own favor (Köbis *et al.*, 2015; Cheeseman and Peiffer, 2020). It is important to be aware of these potential pitfalls of losing citizens' support by overburdening them when implementing AI-ACT. To avoid overwhelming citizens with corruption news, it might thus prove useful to implement AI-ACT in a way that it seeks to minimize wrong accusations (hence particularly reducing false positives) of corruption and potentially provide batched reports of corruption cases rather than a continuous stream of reports.

Further, a successful institutional implementation of AI-ACT requires a proper calibration of the *degree of autonomy* yielded to the algorithms. Here, the academic literature distinguishes broadly between human decision-makers being included in the decision (human-in-the-loop, HITL) or excluded from it (human-out-of-the-loop, HOTL, (Goldenfein, 2019; Starke and Lünich, 2020; Köbis and Mossink, 2021). These varying degrees of autonomy bear importance when using AI-ACT. Consider for example the different implications of whether algorithms merely flag suspicious cases to human investigators, who then work based on such AI reporting. In other instances, a human might not have to actively engage with the case but merely confirm or veto the algorithm's decisions. Then again, in more extreme forms of algorithmic autonomy, AI-ACT disclose suspicious cases without such a human review mechanism. As an example for this seemingly futuristic version of AI-ACT, consider the Brazilian Tweetbot called 'Rosie da Serenata' (see Serenata.ai). It draws on publicly available government data on reimbursement claims of government officials and autonomously detects suspicious cases. It further automatically tweets such cases out to its followers and encourages them to further investigate the cases (Mattoni, 2020).

Given the special importance, let us separately consider the implications of removing humans from the loop in AI-ACT for top-down and bottom-up efforts. First, it is important to note that multiple actors might seek to corrupt the anti-corruption process, in particular within top-down approaches. It is a well-established phenomenon in the corruption literature, that those actors tasked with fighting corruption, like prosecutors or compliance officers, often fall prey to high levels of corruption themselves: a so-called corruption trap emerges (Persson, Rothstein and Teorell, 2013; Fisman and Golden, 2017a; Kubbe and Engelbert, 2017; Stephenson, 2020). Concentrating the authority to make impactful decisions about whom to (not) prosecute in the hands of few further exacerbates this risk. Here, AI-ACT in its most autonomous implementation (HOTL) has unprecedented potential to help escaping this corruption trap. Calibrated correctly, AI-ACT could provide an autonomous, incorruptible agent that makes decisions that (corrupt) human decision makers cannot interfere with.

Conversely, giving final authority about decisions made by AI-ACT to humans (HITL) requires recognizing the human element of such decisions. That is, understanding common human errors, biases and mental shortcuts helps to harness the potential of AI-ACT (Guszcza, 2015). For example, algorithms that are represented via a complicated interface compromise the quality of such human-machine hybrid teams. Further, HITL for top-down efforts requires integrity within the institutions. That is, if decision makers entrusted with using AI-ACT tinker with its decisions to let corrupt culprits off the hook, well-intended anti-corruption efforts run the danger of being rendered obsolete. This danger, however, is less pronounced for bottom-up approaches where those using AI-ACT (e.g. journalists or civil society actors) have overall fewer incentives to withhold or change disclosures of corruption cases in the first place.

Taken together, implementing AI-ACT presents unique pitfalls and potentials for top-down and bottom-up approaches. If AI-ACT are implemented carelessly, they can lead to non-negligible backlash - by public officials who lose trust for being overly surveilled and by journalists and citizens who get an overkill of (inaccurate) corruption news. On the positive side, when implemented carefully, AI-ACT, particularly in top-down approaches, can help to solve one of the long-lasting, intricate challenges for anti-corruption: the corruption trap.

## Future Outlook: Emerging Questions for Society-in-the-Loop

Although considered the next frontier, using AI technologies to curb corruption is still in its early phase of implementation. Therefore, nascent decisions about how to use AI-ACT will shape the way it will affect (future) societies. When designed and implemented right, such tools could help top-down efforts to replace outdated, and corrupted (human) anti-corruption processes with unbiased AI tools of unprecedented speed and computing abilities. For bottom-up efforts, AI algorithms could also be the driving force to mobilize previously apathetic citizens in new efforts to keep power holders accountable. Yet, AI-powered anti-corruption efforts also could go awry. AI tools might just be the next example of innovations that were heralded for their integrity potential, but ended up causing cynicism or, even worse, aggravating existing anti-corruption efforts. A responsible use of AI-ACT needs an active involvement of the citizenry. Put differently, it requires "society in the loop" (Rahwan, 2018). Citizens need to have a say in the way AI-ACT permeates society. We highlight some of the emerging questions for policy-oriented research, that will only become more pressing in the future.

### Input Data - Which data should be off limits?

One trend is for certain: more data will become available. Data that can also be used for AI-ACT. So far, such data stem primarily from (open) government data, leaked data, crowdsourced efforts or (social) media. In the future, new data sources are conceivable. Consider so-called data traces (Flyverbom and Murray, 2018). As digital technologies are increasingly embedded in people's daily lives, human behavior also leaves digital traces that are collected and stored by internet platforms, smartphones, apps, sensors and other devices. Such digital data traces include self-tracking devices, social media communication, geospatial data, browser history and entail contextual data about when, where and how behaviors occur (Rafaeli, Ashtar and Altman, 2019).

Current efforts using AI-ACT are only starting to tap into the sheer unlimited possibilities that such data traces offer. Consider already existing apps that aim to hold powerholders accountable by documenting all interactions between citizens and public officials. In the case of 'Siri, I'm being pulled over' an app automatically records any encounter with police officers (Vincent, 2020). In the near future, such crowdsourced efforts could provide data to train algorithms about which input features help to predict abuse of power, including time of the day, area and even the physical appearance of the police officer. While new types of AI-ACT using such data traces might enter the scene in the near future, there needs to be a broad public discussion about which data sources should be used and which ones should be off limits.

To illustrate the intricacies of such debates, let us zoom in on one particular data source: facial images (Crawford, 2019; de Vries and Schinkel, 2019). For example, the company ClearView, in use in several US police districts has scraped billions of images off the internet from social media platforms such as LinkedIn, Facebook and Twitter (Strong, 2020). Their services help the police to identify suspects. At the same time, experimental research in psychology seems to suggest that unique facial characteristics of "corruptness" are identifiable from a person's face (Lin, Adolphs and Alvarez, 2018). First efforts have combined both streams and built AI models to predict individual-level corruption risks based on facial images, even promising "to predict white collar criminality through real-time facial analysis" (Lavigne, Clifton and Tseng, 2017, p. 8). Not surprisingly, much controversy has emerged around whether, how and where to let AI use facial images for making such predictions (Crawford, 2019).

Although potentially helping to fight the dreadful acts of corruption, would people find the use of such data sources justifiable, even if it leads to privacy breaches? Do citizens' consider the use of data less acceptable when used to *predict* versus *detect* corrupt cases? How does the view of the public differ from the assessment of ethicists, policy makers, and IT developers? Further, who should control and be able to use such data? Is a harder pill to swallow for society when such tools are in the hands of professional prosecutors for internal use in top-down approaches? What if the government is notoriously corrupt itself? Would it then be better off in the hands of journalists, civil society organizations or even average citizens as part of bottom-up efforts?

**Algorithmic Design - How to solve the trade-offs?**

As with all algorithms, AI-ACT face the challenge of trying to achieve many, somewhat incompatible, goals at the same time (Kearns and Roth, 2019). For instance, increasing accuracy (reducing false positive and false negative errors) can come to the cost of explainability (Gunning *et al.*, 2019). That is, sophisticated algorithms drawing on large data sets often produce accurate outcomes that however defy simple explanations - rendering the outcome a "black box" model (Miller, 2019). Those are not the only algorithmic design considerations that can offset each other. Add fairness, privacy and security concerns to the mix and multifaceted trade-offs matrices emerge (Rahwan, 2018).

Understanding citizens' perspectives on the emerging trade-offs in algorithmic design bears immense relevance. How do citizens want developers to resolve such trade-offs? Do they find it more acceptable for the algorithm to err on the conservative side? Is it appropriate to use more accurate algorithms that do however defy simple explanations? When are more "dumbed down", hence explainable but at the same time less accurate algorithms suitable in the fight against corruption? Finally, consider that transparent and explainable algorithms most likely help to build citizens' trust towards the technology, but might also facilitate corrupt actors to game the algorithm.

**Institutional Implementation - How to mobilize and keep citizens involved?**

When implementing AI-ACT a first prerequisite, in particular for bottom-up approaches, consists of mobilizing citizens. The outlined digital crowdsourcing efforts hint at the immense potential, but also provide a warning sign about its shortcomings. On the positive side, crowdsourcing tools can mobilize citizens to report a staggering number of corruption cases. Consider for example, the crowdsourced reporting system Trade Route Incident Mapping System (TRIMS) in Nigeria (giz, 2016). Truckers and small traders reported when and where they were extorted to pay bribes in traffic checkpoints using their mobile phone. On the negative side however, in absence of tangible consequences, the engagement often fizzles and in many cases dies out altogether. It is a common difficulty for crowdsourcing efforts to go beyond the initial mobilization phase and to *keep* people engaged. As a case in point, the website of TRIMS (http://www.trimsonline.org/) is no longer accessible. In addition, reporting on Ipaidabribe has plummeted in recent years.

Could outsourcing several of the labor-intensive tasks to AI help to keep citizens involved? One reason why collective action is hard to sustain stems from the simple fact that it requires effort. It is indeed a lot to ask from an average citizen to research, analyze and communicate often-intricate corruption schemes. Imagine AI-ACT that alleviate some of the burden. In the future, average citizens could consult an app that does not monitor them but instead allows citizens to monitor their government. Citizens could see within a few seconds how political networks and affiliations might have influenced public decisions (Tegmark and Harari, 2019). In a matter of clicks and swipes, citizens could actively contribute to keep public office holders accountable. Eventually, such AI-ACT could facilitate involvement and thus play an unprecedented, democratizing role. Which incentives need to be in place for citizens to use it? Conversely, what psychological barriers might keep them from using it? Could people's willingness to use such an app increase if the code is openly available?

One element that might influence both people's willingness to adopt such AI-ACT, but also their general approval of using AI in the fight against corruption, lies in the degree of human involvement in the decision. Ample research documents that people are generally averse towards algorithms making ethical decisions (Bigman and Gray, 2018; Laakasuo, Köbis and Palomäki, 2021). How much autonomy should be granted to AI and in which sectors? For which areas of anti-corruption should the decision authority remain in the hands of human decision makers? Do citizens' views differ for AI-ACT along top-down versus bottom-up approaches?

To illustrate this point, imagine the use of AI-ACT in the judiciary. Do people find it acceptable if AI-ACT autonomously prevent, detect and even prosecute corruption? Where do citizens draw the line? Already today, algorithms assign judges to crime cases to avoid impartiality and fraud (Algorithm Watch, 2019). Going a step further, what about future scenarios in which AI-ACT do not just assign the cases but make the final decisions in them? Would citizens consider such far-reaching changes to the judicial system appropriate? How does the level of trust towards judicial institutions influence such assessments?

**Conclusion: The importance of gaining empirical answers**

This paper set out from the premise that corruption presents an immense societal challenge around the globe. In light of past failures of anti-corruption efforts, recent advances in AI have ignited new hope. Here, we outlined that AI's autonomous abilities can indeed play an unprecedented role in the fight against corruption, both for top-down and bottom-up approaches. Yet, for both approaches unique potential and pitfalls emerge along the dimensions of input data, algorithmic design and institutional implementation. Harnessing the power of AI-ACT responsibly requires society to be in the loop. Especially, since AI-ACT are still in their initial stage, incorporating citizens' voices marks a key step.

One general way to gain answers to outlined questions consists of conducting research that assesses people's views on AI-ACT. Recent studies using qualitative methods have provided valuable insights into how people respond to algorithmic management, showing that many are averse to such AI managers (Möhlmann and Zalmanson, 2017). Furthermore, large-scale quantitative methods have helped to crowdsource people's moral preferences about how AI technologies - in this case autonomous vehicles - should behave (Awad *et al.*, 2018, 2020). Using similar approaches can help to involve citizens in the development and adoption of AI-ACT. Assessing and recognizing their views and preferences about such tools helps to meet the emerging (ethical) dilemmas that such technological innovations bring about. In particular, since AI-ACT is still in its initial phase such a citizen-centered approach can help to shape its future trajectory, so that AI tools can fulfill some of the hopeful aspirations in the fight against corruption.